\documentclass[12pt]{article}
\usepackage{graphicx}

\def\ignore#1{{}}

\newcounter{sxn}

\newcounter{axn}

\date{}

\newdimen\mybaselineskip
\newcommand{\beeq}{\begin{equation}}
\newcommand{\eneq}{\end{equation}}
\newcommand{\beqn}{\begin{eqnarray}}
\newcommand{\eeqn}{\end{eqnarray}}

\def\la{\raise.16ex\hbox{$\langle$}\lower.16ex\hbox{}  }
\def\ra{\, \raise.16ex\hbox{$\rangle$}\lower.16ex\hbox{} }

\def\Pl{{\rm Pl}}

\def\psibar{ \psi \kern-.65em\raise.6em\hbox{$-$} \lower.6em\hbox{} }
\def\psibarb{ \psi \kern-.65em\raise.6em\hbox{$-$}  }

\begin{document}

\thispagestyle{empty}

\baselineskip=12pt



\vspace*{1.cm}

\begin{center}  
{\LARGE \bf  Asymptotic Spectrum of Kerr Black Holes in the Small Angular Momentum Limit}
\end{center}

\baselineskip=14pt

\vspace{1cm}
\begin{center}
{\bf  Ramin G. Daghigh$\sharp$, Michael D. Green$\dagger$, and Brian W. Mulligan$\ast$}
\end{center}

\vspace{0.25 cm}
\centerline{\small \it $\sharp$ Natural Sciences Department, Metropolitan State University, Saint Paul, Minnesota, USA 55106}
\vskip 0 cm
\centerline{} 

\vspace{0.25 cm}
\centerline{\small \it $\dagger$ Mathematics Department, Metropolitan State University, Saint Paul, Minnesota, USA 55106}
\vskip 0 cm
\centerline{} 
 
\vspace{0.25 cm}
\centerline{\small \it $\ast$ School of Physics and Astronomy, University of Minnesota, Minneapolis, Minnesota, USA 55455}
\vskip 0 cm
\centerline{}

\vspace{0cm}
\begin{abstract}
We study analytically the highly damped quasinormal modes of Kerr black holes in the small angular momentum limit.  To check the previous analytic calculations in the literature, which use a combination of radial and tortoise coordinates, we reproduce all the results using the radial coordinate only.  According to the earlier calculations, the real part of the highly damped quasinormal mode frequency of Kerr black holes approaches zero in the limit where the angular momentum goes to zero.  This result is not consistent with the Schwarzschild limit where the real part of the highly damped quasinormal mode frequency is equal to $c^3 \ln(3)/(8\pi G M)$.  In this paper, our calculations suggest that the highly damped quasinormal modes of Kerr black holes in the zero angular momentum limit 
make a continuous transition from the Kerr value to the Schwarzschild value.  We explore the nature of this transition using a combination of analytical and numerical techniques.  Finally, we calculate the highly damped quasinormal modes of the extremal case in which the topology of Stokes/anti-Stokes lines takes a different form.
\baselineskip=20pt plus 1pt minus 1pt
\end{abstract}

\newpage

\section{Introduction}

\hspace{0.5cm} Black hole quasinormal modes (QNMs) are the natural vibrational modes of perturbations in the spacetime exterior to an event horizon.  QNM frequency spectrum is composed of an infinite number of discrete and complex frequencies, $\omega_n=\omega_R+i\omega_I$, with the ``overtone'' number $n=1,2,3,\dots$.  The imaginary part of the frequency, $\omega_I$, signals the presence of damping, a necessary consequence of boundary conditions that require energy to be carried away from the system.   

Black holes are used for exploring ideas in quantum gravity, and their QNM frequency spectrum is an obvious place to search for a quantum signature.  One possible link between QNMs with high damping rates and the semi-classical level spacing in the black hole quantum area spectrum has been proposed originally by Hod\cite{Hod} in 1998 and modified recently by Maggiore\cite{Maggiore}.  Many people have since calculated the highly damped QNMs of different black hole models.  

The QNMs of Kerr black holes with high damping rates were first explored numerically by Berti et al.\ in \cite{Berti-K, Berti-C-K, Berti-C-Y}.  It was not until recently that Keshet and Hod\cite{Keshet-Hod} were able to analytically calculate the highly damped QNMs of Kerr black holes for the first time.  This work followed with a more detailed paper by Keshet and Neitzke\cite{Keshet-N}.  Kao and Tomino\cite{Kao-T} generalized the results in \cite{Keshet-Hod} to spacetime dimensions greater than four for scalar (spin-zero) perturbations.  According to the results obtained in \cite{Berti-K, Berti-C-K, Berti-C-Y, Keshet-Hod, Keshet-N, Kao-T}, the real part of the highly damped QNM frequencies of Kerr black holes approaches zero in the Schwarzschild limit of $a\rightarrow 0$, where $a$ is the black hole angular momentum per unit mass.  This is in clear contradiction with the Schwarzschild result in which the real part of QNM frequencies approaches $\ln(3)/(8\pi  M)$, where $M$ is the black hole mass, as the damping rate approaches infinity.  (Here, we use geometrized units where $G=c=k_B=1$.)  In other words, it appears that the large damping limit ($|\omega_I|\rightarrow \infty$) and the zero angular momentum limit ($a\rightarrow 0$) of Kerr QNMs do not commute.  In \cite{Keshet-Hod}, Keshet and Hod state that ``the asymptotic QNMs are not continuous at $a=0$".  In other words, at $a=0$ the real part of the QNM frequency makes a discontinuous transition from zero to the Schwarzschild value of $\ln(3)/(8\pi  M)$.  This issue was presented by Maggiore in \cite{Maggiore} as an argument against Hod's proposal in \cite{Hod} and consequently provides one of the motivations for this work.  Note that the imaginary part of the asymptotic Kerr QNMs does not suffer from such discontinuity. 

Musiri and Siopsis\cite{Musiri-S} show analytically that when the range of QNM frequencies is bounded from above by $1/a$ ($|a\omega| < 1$), the Kerr QNMs do coincide with the Schwarzschild QNMs in the high damping limit.  The calculation in  \cite{Musiri-S} is valid for small values of the parameter $a$ and includes the Schwarzschild case where $a=0$.  According to \cite{Musiri-S}, in the large damping limit 
\beeq
\omega_R \rightarrow {{\ln (3)}\over {8\pi M}} +m a ~,
\label{asymptotic-omega}
\eneq 
where $m$ is the azimuthal eigenvalue of the wave.  

In this paper we use a different analytic technique to reproduce the result obtained by Musiri and Siopsis in \cite{Musiri-S}.  We also reproduce the result obtained by Keshet and Hod in \cite{Keshet-Hod} using a slightly different analytic technique in which the tortoise coordinate is not required.  We show that the result obtained in \cite{Keshet-Hod, Keshet-N} is only valid when $|a\omega| > 1$.  
We also show that for any large but finite QNM frequency, the transition from the Kerr value to the Schwarzschild value happens in a continuous way.  We explore in detail this transition which happens at around $|a\omega| \approx 1$.  Such transition becomes discontinuous only when $|\omega|=\infty$.  In this paper, we also calculate the highly damped QNMs in the interesting case of extremal Kerr black holes for the first time.  

In section 2, we set up the problem. We present the analytic calculations in section 3.  In section 4, we explain the numerical calculations and provide the results.  We conclude the paper in section 5.

\section{Wave Equation}

\hspace{0.5cm}Following the notations used by Keshet and Hod in \cite{Keshet-Hod}, the radial part of Teukolsky's wave equation for a Kerr black hole with mass $M$ and angular momentum $J$ can be written in the general form 
\beeq
{d^2\Psi(r) \over dr^2}+P(r)\Psi(r) =0 ~,
\label{Wave-Eq}
\eneq 
where
\beeq
P(r)= {{q_0(r)\omega^2+q_1(r)\omega +q_2(r)}\over \Delta^2 }~.
\label{R-function}
\eneq 
We define
\beeq
\Delta=r^2-2Mr+a^2+q^2~,
\label{Delta}
\eneq 
\beeq
q_0 \equiv (r^2+a^2)^2-a^2\Delta~,
\label{q0}
\eneq 
\beeq
q_1 \equiv -2am(2Mr-q^2)+2is\left[r(\Delta+q^2)-M(r^2-a^2)\right]~,
\label{q1}
\eneq 
\beeq
q_2 \equiv m^2a^2-(A_{lm}+s) \Delta+M^2-a^2-q^2-s(M-r)\left[2iam+s(M-r)\right]~,
\label{q2}
\eneq 
where $a=J/M$, $l$ and $m$ are angular and azimuthal harmonic indices, $A_{lm}$ is the separation constant, and $s=0,-1/2, -1, -2$ is the spin-weight parameter for scalar, two-component neutrino, electromagnetic, and gravitational fields respectively.  Note that Teukolsky's wave equation can be generalized to include black holes with electric charge $q$\cite{Dudley}, where the case of $q\neq 0$ in the above equations is understood only for scalar fields.  The roots of $\Delta$ determine the radii of outer and inner horizons.  These roots are $r_{\pm}=M\pm\sqrt{M^2-a^2-q^2}$.  

The differential equation ({\ref{Wave-Eq}) has two linearly independent WKB solutions:
\beeq
\left\{ \begin{array}{ll}
                   f_{1}^{(t)}(r)=Q^{-1/2}(r)e^{+i\int_{t}^rQ(r')dr'}~\\
                   \\
                   f_{2}^{(t)}(r)=Q^{-1/2}(r)e^{-i\int_{t}^rQ(r')dr'}~
                   \end{array}
           \right.        
\label{WKB}
\eneq
where 
\beeq
Q^2(r)=P(r) - {1 \over 4(r-r_-)^2}-{1 \over 4(r-r_+)^2}~.
\label{Q-P}
\eneq 
The last two terms on the right hand side of the above equation come from matching the WKB solutions to the exact solutions of the wave equation (\ref{Wave-Eq}) in the limit where $r\rightarrow r_\pm$.

The boundary conditions that we apply to the wave equation (\ref{Wave-Eq}) are outgoing-wave at infinity and ingoing-wave at the horizon.  We will choose the phase of the square-root of $Q^2$ such that $\sqrt{Q^2} \sim \omega$ as $r\rightarrow \infty$.  This means that the outgoing-wave solution at infinity is proportional to $f_1$ while the ingoing-wave solution at $r_+$ is proportional to $f_2$.  Here, we assume that perturbations depend on time as $e^{-i\omega t}$.  This means that $\omega_I <0$.

In the large damping limit ($\omega_I\rightarrow -\infty$), the angular separation constant can be approximated to be 
\beeq
A_{lm}=iA_1a\omega+(A_0+m^2)+O(|\omega|^{-1})~
\label{Alm}
\eneq 
and in the Schwarzschild limit where $a\rightarrow 0$, this constant can be approximated to be
\beeq
A_{lm}=l(l+1)-s(s+1)+O(a\omega)~.
\label{Alm-intermediate}
\eneq

It is easy to show that in the limit $a \rightarrow 0$, and as long as $|\omega|  \ll 1/a$, we get  
\begin{eqnarray}
q_0 &\approx& r^4, \nonumber \\
q_1&\approx&2is(r^3-3Mr^2),  \nonumber \\
q_2&\approx&-A_{lm}(r^2-2Mr)+{{1-s^2}\over 4}(4r^2+4M^2-8rM)-(1-s)(r^2-2Mr)~.
\end{eqnarray}
If, in addition to the above conditions, we assume that $|\omega| \gg 1$, we can simplify $q_0$, $q_1$, and $q_2$ in the expression $q_0\omega^2+q_1\omega+q_2$ further to find that
\begin{eqnarray}
q_0&\approx&r^4~,  \nonumber \\
q_1&\approx&-6isMr^2~,  \nonumber \\
q_2&\approx&{{1-s^2}\over 4}(2M)^2~. 
\label{zero-a-qs}
\end{eqnarray}
Note that we cannot ignore the term $q_1\omega$ because this term becomes comparable to $q_0 \omega^2$ and $q_2$ when $r$ is of the order of magnitude of $|\omega|^{-1/2}$.

\section{Analytic Calculations}

\hspace{0.5cm}In order to extract the WKB condition on the asymptotic QNM frequencies, first, we need to determine the zeros and poles of the function $Q^2$ and consequently the behavior of the Stokes and anti-Stokes lines in the complex $r$-plane.  Stokes lines are the lines on which 
\beeq
\mbox{Re}\int_t^r Q(r')dr'=0~,
\label{Stokes}
\eneq   
and anti-Stokes lines are the lines on which 
\beeq
\mbox{Im}\int_t^r Q(r')dr'=0~,
\label{anti-Stokes}
\eneq  
where $t$ is a zero of the function $Q^2$.  The poles are located at $r_\pm$.

\begin{figure}[tb]
\begin{center}
\includegraphics[height=7cm]{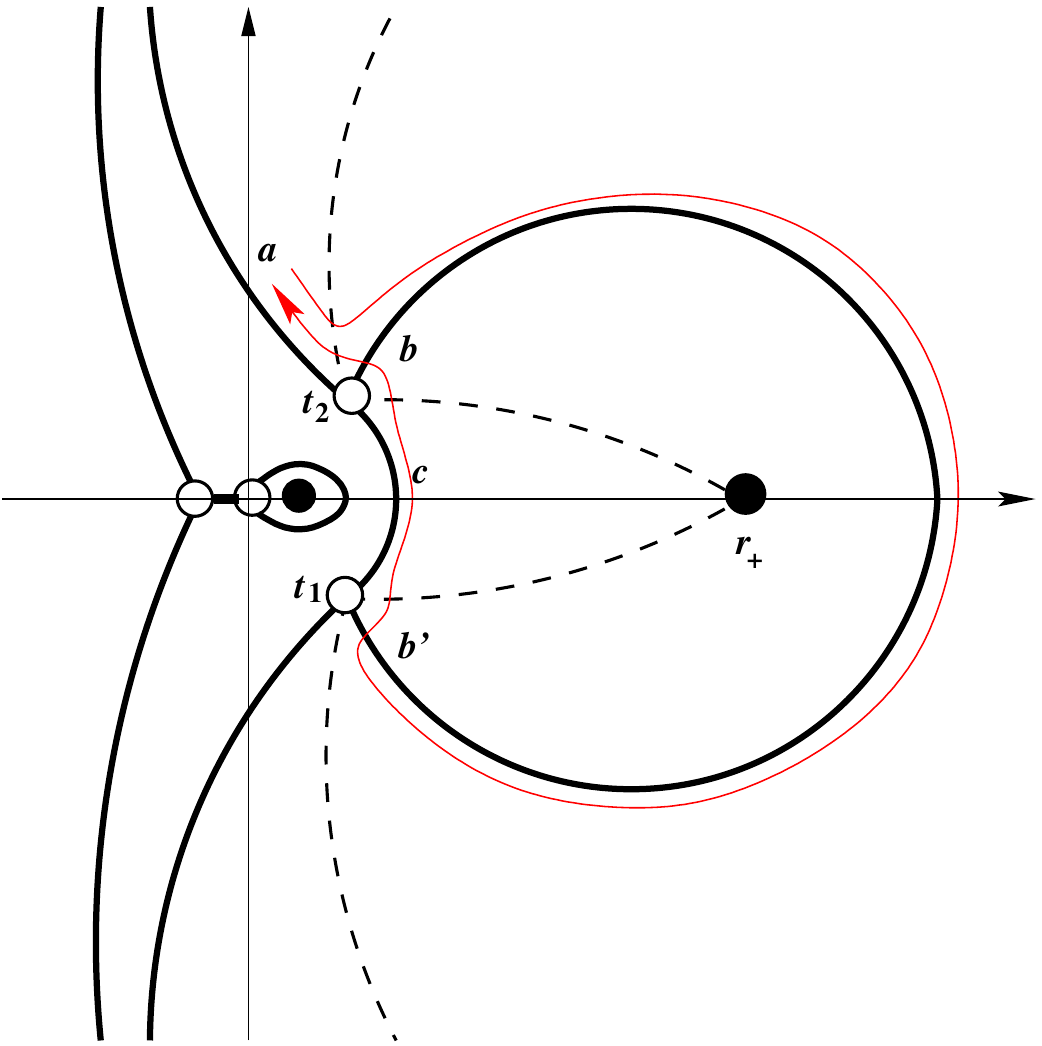}
\end{center}
\caption{\footnotesize A schematic illustration of Stokes (dashed) and anti-Stokes (solid) lines for finite $a$.  The hollow circles represent the zeros of $Q^2$, while the filled circles are the two poles.  The thin line is the path taken along the anti-Stokes lines to derive the WKB condition on $\omega$.}
\label{Fig-a=finite}
\end{figure}

\subsection{The case of finite $a$:} 

\hspace{0.5cm}In this case the zeros of the function $Q^2$ are, by a factor of $|\omega|^{-1}$, close to the zeros of the function $q_0$.  After determining the topology of the Stokes/anti-Stokes lines, we follow the path shown in Fig.\ \ref{Fig-a=finite} starting on an unbounded anti-Stokes line at $a$.  The solution on line $a$ is known due to the boundary condition at infinity:
\beeq
\Psi_{a}=f_1^{(t_2)}~.
\label{psi-a}
\eneq 
We apply the rules that are explained by Andersson and Howls in \cite{Andersson} to move along the anti-Stokes lines.  The steps we take are as follows:
\beeq
\Psi_b = f_1^{(t_2)} - if_2^{(t_2)}  
\eneq
\beeq
\Psi_{b'} = e^{i\tilde{\gamma}_{21}} f_1^{(t_1)} - ie^{-i\tilde{\gamma}_{21}} f_2^{(t_1)}
\eneq
\begin{eqnarray}
\Psi_{c} &=&  \left(e^{i\tilde{\gamma}_{21}} + e^{-i\tilde{\gamma}_{21}} \right) f_1^{(t_1)} - ie^{-i\tilde{\gamma}_{21}} f_2^{(t_1)}  \nonumber \\
&=& \left( e^{i\tilde{\gamma}_{21}}+e^{-i\tilde{\gamma}_{21}}\right) e^{i\gamma_{12}}f_1^{(t_2)} - ie^{-i\tilde{\gamma}_{21}} e^{-i\gamma_{12}}f_2^{(t_2)}
\label{psi-c}
\end{eqnarray}
\beeq 
\Psi_{\bar b} = \left[ \left(e^{i\tilde{\gamma}_{21}} + e^{-i\tilde{\gamma}_{21}} \right) e^{i\gamma_{12}} +e^{-i\tilde{\gamma}_{21}}e^{-i\gamma_{12}}\right] f_1^{(t_2)} - i e^{-i\tilde{\gamma}_{21}}e^{-i\gamma_{12}}   f_2^{(t_2)} 
\eneq
\begin{eqnarray}
\Psi_{\bar{a}} & = & \left[ \left(e^{i\tilde{\gamma}_{21}} + e^{-i\tilde{\gamma}_{21}} \right) e^{i\gamma_{12}} +e^{-i\tilde{\gamma}_{21}}e^{-i\gamma_{12}}\right] f_1^{(t_2)}   \nonumber \\
 & &  
  + i \left(e^{i\tilde{\gamma}_{21}} + e^{-i\tilde{\gamma}_{21}}\right) e^{i\gamma_{12}}  f_2^{(t_2)}~
\end{eqnarray}
where
\beeq
\gamma_{12} =-\gamma_{21}=\int_{t_1}^{t_2} Q dr ~
\eneq
along a path to the left of the event horizon and
\beeq
\tilde{\gamma}_{12} = \int_{t_1}^{t_2} Q dr ~
\eneq
along a path to the right of the event horizon.  The boundary condition at infinity requires that the coefficient of $f_2$ in $\Psi_{\bar{a}}$ be equal to zero.  This leads to the WKB condition 
\beeq
e^{2i\tilde{\gamma}_{12}}=-1~.
\label{WKB-finite-a}
\eneq
Note that the above condition gives us the correct monodromy as a result of rotating clockwise around the outer horizon in the complex $r$-plane, which is
\beeq
\Psi_{\bar a}=e^{-i\Gamma}\Psi_a~,
\eneq
where
\beeq
\Gamma=\gamma_{12}+\tilde{\gamma}_{21}=\mathop{\oint}_{clockwise} Q dr~.
\eneq
In the large $\omega$ limit, we can write
\beeq
2i\tilde{\gamma}_{12}\approx 2i \omega \int_{t_1}^{t_2} {{\sqrt{q_0}}\over \Delta} dr +2i \int_{t_1}^{t_2} {q_1 \over {2 \sqrt{q_0} \Delta}} dr = \omega \delta_0+is\delta_s+iA_1\delta_A+m\delta_m~,
\eneq
where 
\begin{eqnarray}
\delta_0  &=& 2i \int_{t_1}^{t_2} {{\sqrt{q_0}}\over \Delta} dr,~\delta_m=2i \int_{t_1}^{t_2} {-a(2Mr-Q^2)\over \sqrt{q_0}\Delta} dr,~ \nonumber \\
\delta_A &=& 2i\int_{t_1}^{t_2} {-a\over 2\sqrt{q_0}} dr,   ~ \mbox{and} ~\delta_s= 2i\int_{t_1}^{t_2} {r(\Delta+Q^2)-M(r^2-a^2) \over \sqrt{q_0}\Delta } dr~.
\end{eqnarray}
It is now easy to show that the WKB condition (\ref{WKB-finite-a}) results in 
\beeq
\omega = -m \hat{\omega}-i(\hat{\phi}-n\hat{\delta})~,
\label{omega-finite-a}
\eneq
where $n$ is a large integer and $\hat{\omega}=\delta_m/\delta_0$, $\hat{\delta}=2\pi/\delta_0$, and $\hat{\phi}=(s\delta_s+A_1 \delta_A - \pi)/\delta_0$.  Equation (\ref{omega-finite-a}) is the same equation that Keshet and Hod found in \cite{Keshet-Hod}.

\subsection{The case of $a\rightarrow 0$:} 

\hspace{0.5cm}In the limit $a\rightarrow 0$, as long as $|\omega|>>1$ and $|\omega a|<<1$, we can combine equations (\ref{R-function}), (\ref{Q-P}), and (\ref{zero-a-qs}) to see that the four zeros of the function $Q^2$ approach 
\beeq
\pm\sqrt{i{(3\pm 2\sqrt{2})Ms \over \omega}}~.
\eneq  
Since $s<0$ for all types of perturbations and $\omega \approx -i|\omega|$, the four zeros above are located on the real axis.  The structure of Stokes and anti-Stokes lines are shown in Fig.\ \ref{Fig-a=0}.  In order to derive a WKB condition on the QNM frequency $\omega$ we will follow the path indicated by the thin line shown in Fig.\ \ref{Fig-a=0} along anti-Stokes lines.  We begin our path on the anti-Stokes line labeled $a$ which extends to infinity.  According to the boundary condition, the solution on this line is 
\beeq
\Psi_{a}=f_1^{(t_1)}~.
\eneq 
\begin{figure}[tb]
\begin{center}
\includegraphics[height=5.5cm]{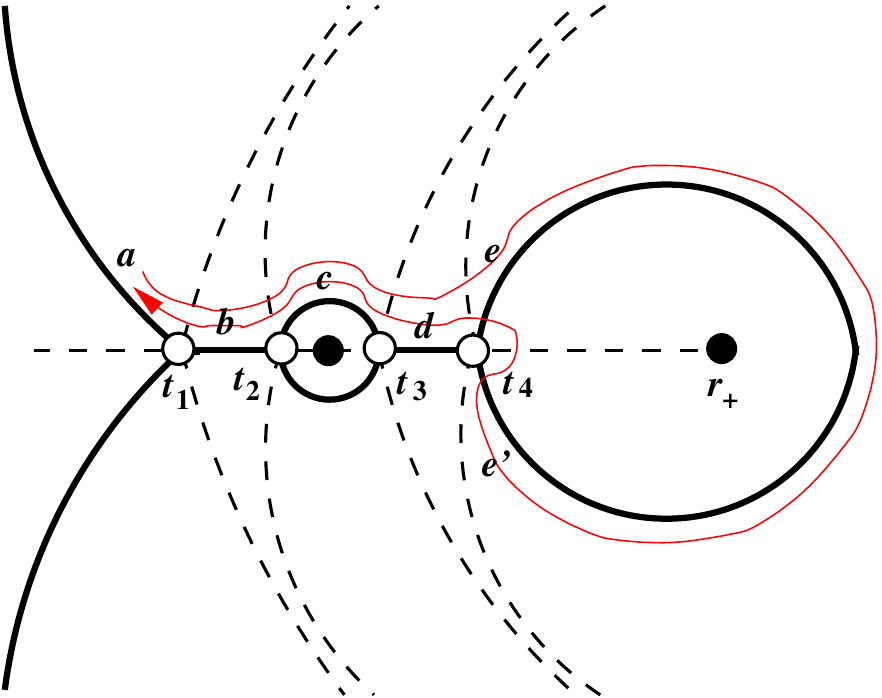}
\end{center}
\caption{\footnotesize A schematic illustration of Stokes (dashed) and anti-Stokes (solid) lines when $a \rightarrow 0$.  The hollow circles represent the zeros of $Q^2$, while the filled circles are the two poles.  The thin line is the path taken to derive the WKB condition on $\omega$.}
\label{Fig-a=0}
\end{figure}
The steps that we take are very similar to the the steps explained in the previous subsection.  We choose our branch cuts in a way that the WKB solution $f_1$ is dominant on the Stokes lines extending to infinity and $f_2$ is dominant on the Stokes line ending at the outer horizon.  After making a complete loop around the horizon, we return to the anti-Stokes line $a$.  The final solution that we get is 
\begin{eqnarray}
\Psi_{\bar{a}} & = & e^{-i \tilde{\gamma} _{44}} \left(1+e^{-2 i \gamma _{12}}e^{-2 i \gamma _{23}}e^{-2 i \gamma _{34}}+e^{-2 i \gamma _{23}}e^{-2 i \gamma _{34}}+e^{-2 i \gamma _{34}}+e^{2 i \tilde{\gamma} _{44}}\right)
f_1^{(t_1)}  \nonumber \\
  & &  
  +i   e^{-i \tilde{\gamma} _{44}} \left(1+e^{2 i \gamma _{12}}+e^{2 i \gamma _{12}}e^{2 i \gamma _{23}}+e^{2 i \gamma _{12}}e^{2 i \gamma _{23}}e^{2 i \gamma _{34}}\right) \nonumber \\
 & & \left(e^{-2 i \gamma _{12}}e^{-2 i \gamma _{23}}e^{-2 i \gamma _{34}}+e^{-2 i \gamma _{23}}e^{-2 i \gamma _{34}}+e^{-2 i \gamma _{34}}+e^{2 i \tilde{\gamma} _{44}}\right)f_2^{(t_1)}~.
\label{end-a=0}
\end{eqnarray}
The boundary condition at infinity requires that the coefficient of $f_2$ in solution (\ref{end-a=0}) be equal to zero.  This leads us to the WKB condition
\beeq
e^{2i\tilde{\gamma}_{44}}=-e^{-2i{\gamma}_{12}}e^{-2i{\gamma}_{23}}e^{-2i{\gamma}_{34}}-e^{-2i{\gamma}_{23}}e^{-2i{\gamma}_{34}}-e^{-2i{\gamma}_{34}}~,
\label{WKB-Schwarzschild}
\eneq
where
\beeq
\tilde{\gamma}_{44}=\mathop{\oint}_{clockwise} Q dr= -2\pi i \mathop{Res}_{r=r_+} Q= -4\pi i\omega M ~
\label{Gamma}
\eneq
and in the large damping limit
\beeq
{\gamma}_{ij}\approx \int_{t_i}^{t_j}  {r \over 2M}\left(\omega^2-{6isM\omega \over r^2}-{s^2 M^2 \over r^4}\right)^{1/2}dr~.
\label{gamma-a0}
\eneq
Note that condition (\ref{WKB-Schwarzschild}) gives us the correct monodromy as a result of rotating around the outer horizon in the clockwise direction, which is 
\beeq
\Psi_{\bar a}=e^{-i\tilde{\gamma}_{44}}\Psi_a~.
\eneq

\begin{figure}[tb]
\begin{center}
\includegraphics[height=7cm]{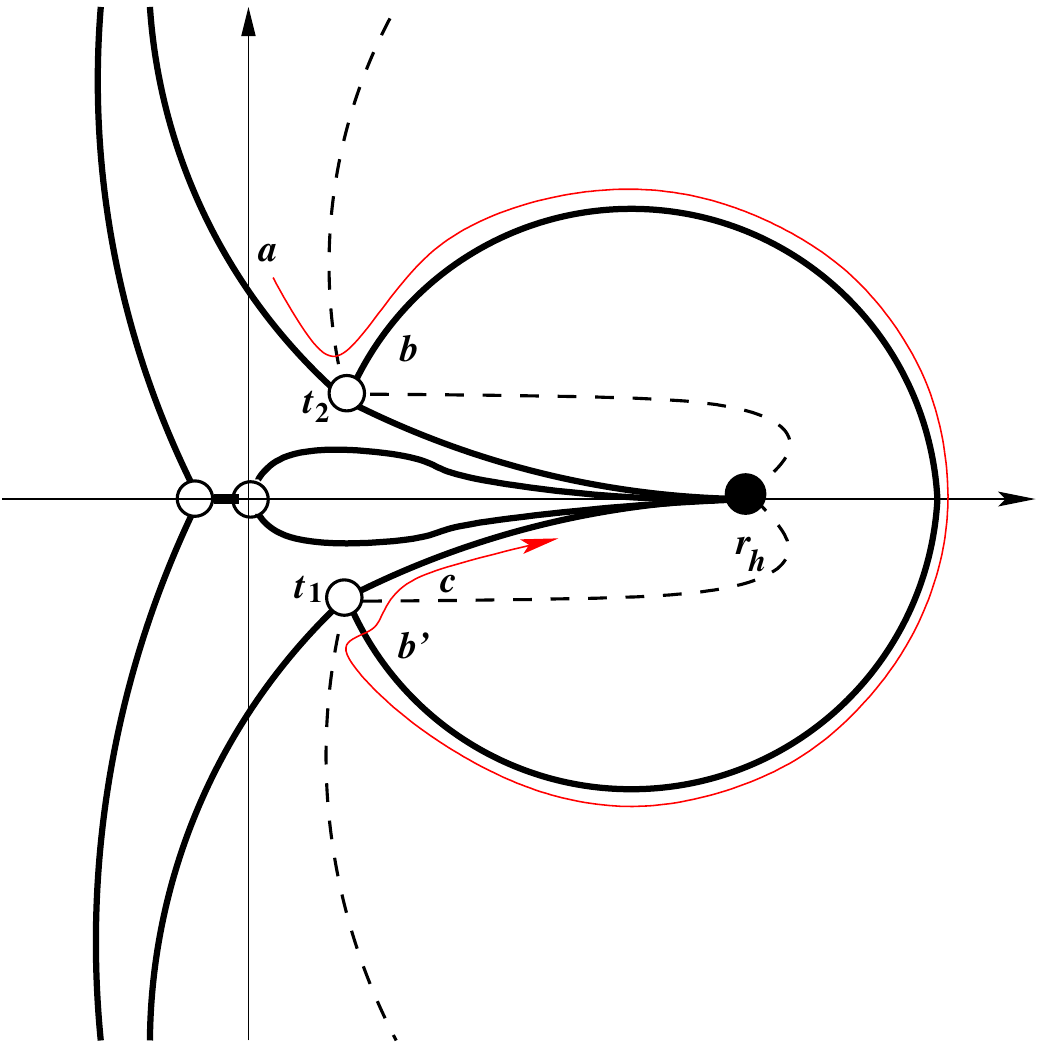}
\end{center}
\caption{\footnotesize A schematic illustration of Stokes (dashed) and anti-Stokes (solid) lines for the extremal case.  The hollow circles represent the zeros of $Q^2$, while the filled circle is the pole at the point where the inner and outer horizons coincide.  The thin line is the path taken along the anti-Stokes lines to derive the WKB condition on $\omega$.}
\label{Fig-extremal}
\end{figure}
\begin{figure}[tb]
\begin{center}
\includegraphics[height=7cm]{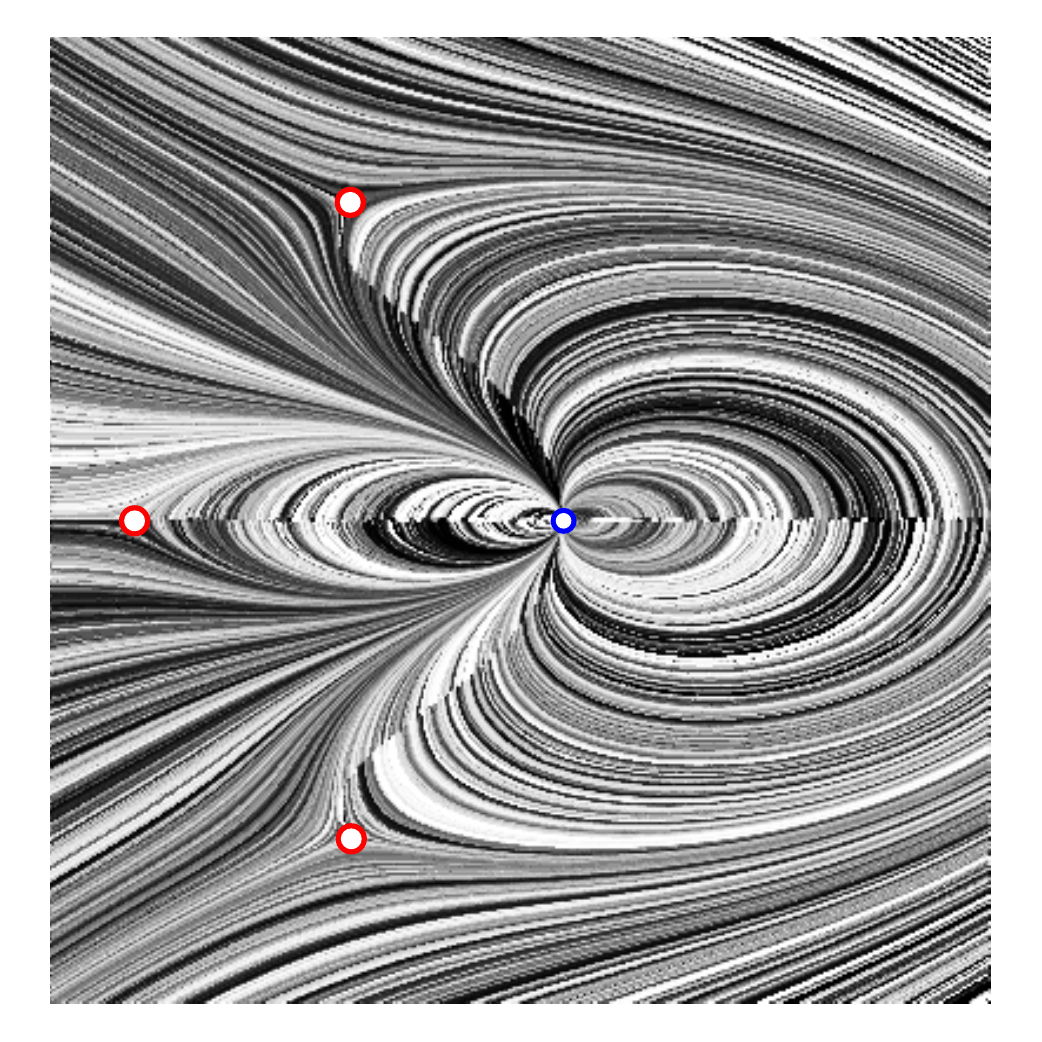}
\includegraphics[height=7cm]{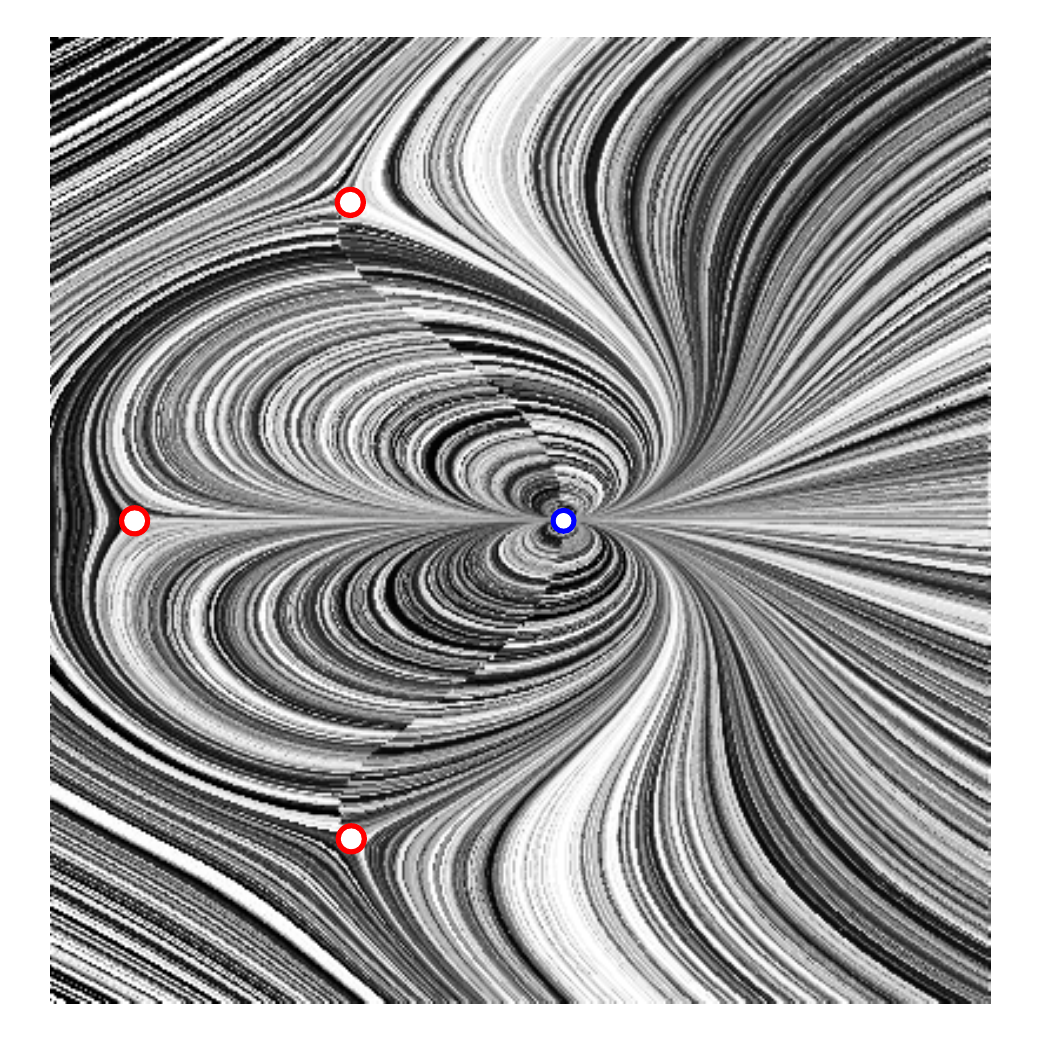}
\end{center}
\caption{\footnotesize Numerically generated anti-Stokes (left) and Stokes (right) lines for the extremal case.  The small hollow circle in the middle is the pole at the inner horizon and the bigger hollow circles are the location of the zeros of $Q^2$. }
\label{Fig-NUM-extremal}
\end{figure}
\begin{figure}[tb]
\begin{center}
\includegraphics[height=4.7cm]{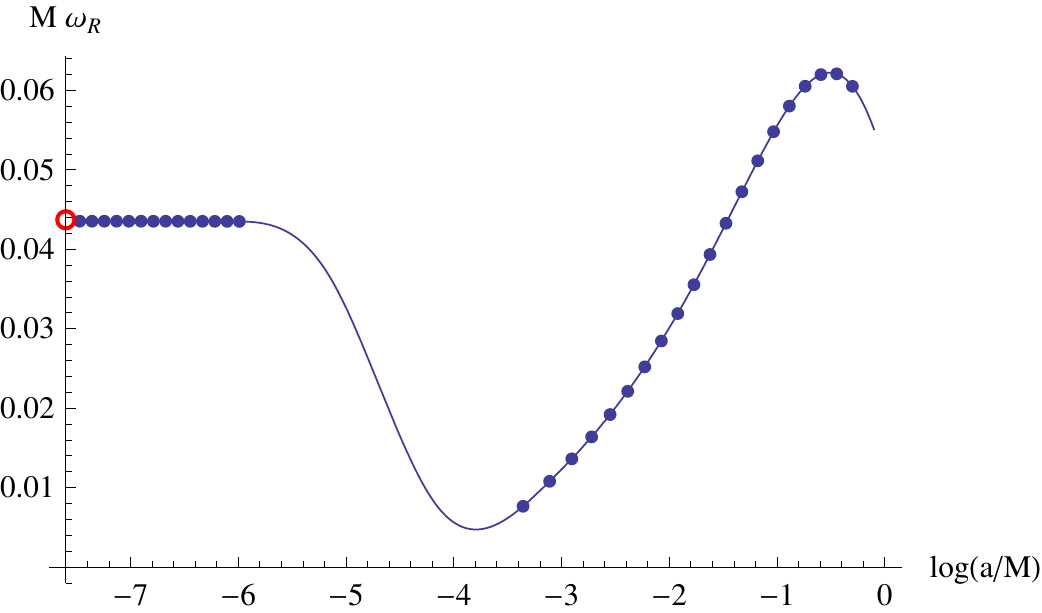}
\includegraphics[height=4.7cm]{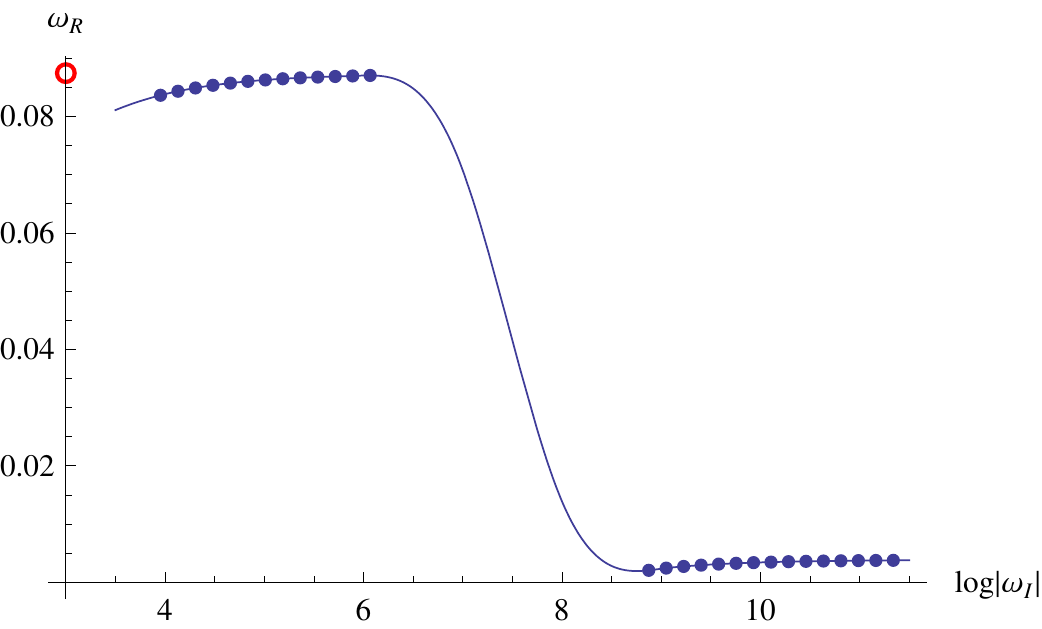}
\end{center}
\caption{\footnotesize The real part of the highly damped QNM frequency as a function of $a$ (left) and the damping rate (right).  Logarithmic scale is used along the horizontal axis to better illustrate the transition region from the Kerr value to the Schwarzschild value. The dots are the numerically generated data points and the solid line is a cubic spline displayed to suggest the possible behavior in the transition region.  The hollow circle on the vertical axis indicates the location of the Schwarzschild value $\omega_R=\ln(3)/8\pi M$.  Note that $\omega_R$ does not appear to be discontinuous at $a=0$. We take $M=1/2$, $m=-1$, $\omega_I=-10^6$ (left), and $a=10^{-6}$ (right).}
\label{Fig-trans}
\end{figure}
\begin{figure}[tb]
\begin{center}
\includegraphics[height=9cm]{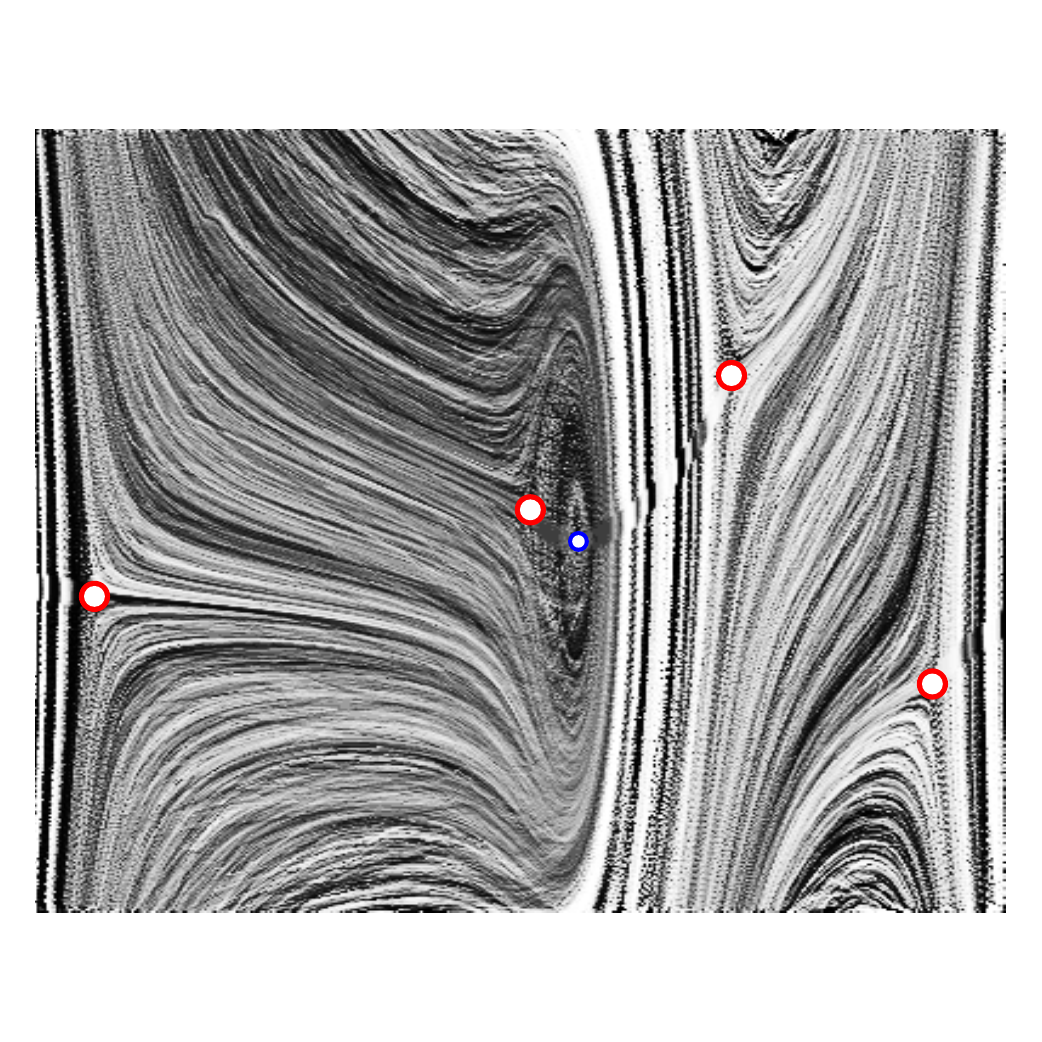}
\end{center}
\caption{\footnotesize Numerically generated anti-Stokes lines in the transition region from the Kerr topology to the Schwarzschild topology.  The small hollow circle in the middle is the pole at the inner horizon and the bigger hollow circles are the location of the zeros of $Q^2$. }
\label{Fig-NUM-trans}
\end{figure}

The integral (\ref{gamma-a0}) can be solved analytically.  We use the change of variable $y=\omega r^2/sM$ to get
\begin{eqnarray}
\gamma_{ij}&=& {s\over 4}\int^{(3\pm 2\sqrt{2})i}_{(3\mp 2\sqrt{2})i}\left(1-{6i \over y}-{1\over y^2}\right)^{1/2}dy \nonumber \\
&=& \left[ \sqrt{y^2-6iy-1}-\mbox{arccot}\left(3y-\sqrt{y^2-6iy-1}\right)-3i\mbox{arcsinh} \left({-3i+y \over 2\sqrt{2}}\right) \right.\nonumber \\
&& \left. +{i \over 2}\ln\left({y\over 3i-5y+3\sqrt{y^2-6iy-1}}\right)\right]_{(3\mp 2\sqrt{2}) i}^{(3\pm 2\sqrt{2})i} =\mp {s\over 2} \pi~.
\label{gamma-q=0}
\end{eqnarray}
Combining Eqs.\ (\ref{WKB-Schwarzschild}) and (\ref{gamma-q=0}), we find
\beeq
e^{2i\tilde{\gamma}_{44}}= -3~
\eneq
for gravitational and scalar perturbations and
\beeq
e^{2i\tilde{\gamma}_{44}}= -1~
\eneq
for electromagnetic perturbations.  These WKB conditions are in perfect agreement with the Schwarzschild case.

\subsection{The extremal case:} 
\hspace{0.5cm}In the case of extremal Kerr black hole spacetime where the inner and outer horizons coincide ($a=M$), the topology of Stokes/anti-Stokes lines takes a new form shown in Fig.\ \ref{Fig-extremal}.  In order to determine the WKB condition, we follow the path shown in Fig.\ \ref{Fig-extremal}, which begins on the anti-Stokes line labeled $a$ that extends to infinity and after moving from $t_2$ to $t_1$ along the anti-Stokes line to the right of the horizon the path ends on an anti-Stokes line which connects to the horizon $r_h$.  The steps that we take will be identical to the ones we took in subsection 3.1 from Eq.\ (\ref{psi-a}) to (\ref{psi-c}).  The boundary condition at the horizon requires that the coefficient of $f_1$ in Eq.\ (\ref{psi-c}) be equal to zero.  This will lead us to the same WKB condition that we found in Eq.\ ({\ref{WKB-finite-a}).  This shows that the highly damped QNMs of Kerr black holes in the limit $a \rightarrow M$ does indeed coincide with the highly damped QNMs of extremal Kerr black holes, even though the topology of Stokes/anti-Stokes lines in the extremal case is different.  In other words, to find the asymptotic QNM frequency of extremal Kerr black hole all we need to do is to replace the angular momentum $a$ with mass $M$ in Eq.\ ({\ref{omega-finite-a}).  This result is consistent with the high overtone QNMs of other types of extremal black holes\cite{Daghigh-Green}.

\section{Numerical Results}

The configuration of the anti-Stokes lines was found using the LineIntegralConvolutionPlot
in Mathematica applied to the vector field $\langle \cos[\alpha(r)],-\sin[\alpha(r)]\rangle $ where $\alpha(r) =
\arg[Q(r)]$. This vector field defines a direction for $dr$ at every point in the complex plane
such that $\arg[Q(r)dr] = 0$. The anti-Stokes lines follow this vector field to and from the zeros. Similarly, the Stokes lines follow the vector field $\langle \sin[\alpha(r)],\cos[\alpha(r)]\rangle $. Fig.\ \ref{Fig-NUM-extremal}
shows the vector fields for the anti-Stokes and Stokes lines for the extremal case illustrated
schematically in Fig.\ \ref{Fig-extremal}.

The numerical results in Fig.\ \ref{Fig-trans} use numerical integration to find the integrals $\gamma_{ij} = \int_{t_i}^{t_j} Q(r) dr$ along a path in the complex plane homotopic to the anti-Stokes lines.  We then use the WKB conditions (\ref{WKB-finite-a}) and (\ref{WKB-Schwarzschild}) to calculate $\omega_R$ in different regions.  Thus the numerical techniques, as well as the analytical techniques, used in this paper are not valid when the anti-Stokes lines do not connect the zeros of $Q(r)$.  As the zeros move from the configuration seen in Fig.\ \ref{Fig-a=finite} to that seen in Fig.\ \ref{Fig-a=0} the anti-Stokes lines no longer connect the zeros as can be seen by examining the vector field in Fig.\ \ref{Fig-NUM-trans}.  

The results in Fig.\ \ref{Fig-trans} show that when $a \ll {1\over |\omega _I|}$, $\omega_R \rightarrow \ln(3)/(8\pi M)$ and for $a \gg {1\over |\omega _I|}$, the results agree with the numerical and analytical results in \cite{Berti-K, Berti-C-K, Berti-C-Y, Keshet-Hod, Keshet-N, Kao-T}.

\section{Conclusions}

\hspace{0.5cm}In this paper, we used a different analytic technique to reproduce the result obtained by Musiri and Siopsis in \cite{Musiri-S}.  In addition, in order to check the previous analytic calculations of the highly damped QNMs of Kerr black holes\cite{Keshet-Hod, Keshet-N}, we reproduced the results without the use of the tortoise coordinate.  Our calculations suggest that, contrary to the previous results in the literature, the highly damped QNMs of Kerr black holes in the zero angular momentum limit make a continuous transition to the Schwarzschild value.  See Fig.\ \ref{Fig-trans}.  A similar situation exists also in the highly damped QNMs of Reissner-Nordstr$\ddot{\mbox{o}}$m black holes.  In that case as well, it was initially thought that if
we first consider the asymptotic QNMs of a Reissner-Nordstr$\ddot{\mbox{o}}$m black hole and
then let the charge go to zero, $\omega_R$ does not reduce to $\ln(3)/(8 \pi M)$, but
rather reduces to $\ln(5)/(8 \pi M)$\cite{Motl2}.  Such discrepancies in Kerr and Reissner-Nordstr$\ddot{\mbox{o}}$m black holes were used by Maggiore\cite{Maggiore} as an argument against the validity of Hod's conjecture\cite{Hod}. However, similar to what we did here for Kerr black holes, it is shown in \cite{Dagigh-K-O} that the highly damped QNMs of Reissner-Nordstr$\ddot{\mbox{o}}$m black holes in the zero charge limit do coincide with the known Schwarzschild results.  

If we try to derive the horizon area spectrum of a slowly rotating black hole assuming that the real part of the highly damped QNM frequency approaches zero when $a\rightarrow 0$, we will end up with zero spacing between the consecutive states of the spectrum according to Hod's interpretation of the highly damped QNMs\cite{Hod}.   For that reason, Vagenas\cite{Vagenas} and Medved\cite{Medved} use the imaginary part of the QNM frequency, which approaches the Schwarzschild value when $a\rightarrow 0$, in deriving the area spectrum of slowly rotating black holes following Maggiore's interpretation\cite{Maggiore}.  The results of this paper show that Hod's interpretation also gives a non-zero spacing between the consecutive states of the area spectrum.  Following the steps of Medved in \cite{Medved}, we can replace $\omega$ in the adiabatic invariant quantity
\beeq
\int {{dM-\Omega dJ}\over \omega}~,
\eneq
where $\Omega=a/(r_+^2+a^2)$, with $\ln(3)/(8\pi M)$ and we obtain an area spectrum for slowly rotating black holes of the form
\beeq
A_k+\mathcal{O}(J_k^4)\approx 4k\ln(3)l_{Pl}^2~,
\eneq
where $k$ is a large integer.  In the case of finite $a$ neither Hod's interpretation nor Maggiore's interpretation of the highly damped QNMs give a unique value for the spacing in the area spectrum because both $\omega_R$ and the spacing in $\omega_I$ are dependent on the parameter $a$.

It is relevant to mention that Babb et al.\ in \cite{DKB} show that, when one includes quantum corrections to a Schwarzschild spacetime, $\ln(3)$ does not appear in the region of the QNM spectrum where the damping is larger than the
inverse Planck/polymerization length scale, but it does appear in lower damping
rates.  They conclude that it is still plausible to think that the QNMs with damping rates that are large relative
to the inverse horizon length ($\approx M^{-1}$) but small compared to other inverse lengths
will provide information about the black hole horizon.  A similar conclusion can be made regarding the Kerr case.  The QNMs with a damping rate much larger than $M^{-1}$ but much smaller than $a^{-1}$ may provide information about the mass of the black hole only.  The information on the angular momentum of the black hole appears in the QNM spectrum of slowly rotating black holes when the damping rate becomes comparable to $a^{-1}$.  In other words, it seems that for Hod's conjecture to be true, we need to associate $\ln(3)$ with the mass of the black hole only.  When $a$ becomes comparable to the mass of the Kerr black hole, the two regions of the QNM spectrum overlap which makes it difficult to separate the information regarding the mass and angular momentum.  
Now, the question is, how can we relate the QNM frequencies in the intermediate damping region $M^{-1}\ll |\omega_I|\ll a^{-1}$ to the area spectrum of Kerr black holes if they only contain information on $M$.  
One way is to follow the intriguing suggestion of M$\ddot{\mbox{a}}$kel$\ddot{\mbox{a}}$ et al.\ in \cite{Makela1, Makela2} where the authors argue that in the case of multi-horizon black holes, it is the area of both inner and outer horizons which should be quantized in equal steps.  For the Kerr case this area is 
\beeq
A_{total}= 4\pi(r_+^2+a^2)+4\pi(r_-^2+a^2)=16\pi M^2~,
\eneq
which is the same as the area of a Schwarzschild black hole with mass $M$.  We now can use the value of $\omega_R$ in the intermediate damping region of the Kerr QNM spectrum to find that 
\beeq
\Delta A_{total}= 32\pi M \Delta M=32\pi M  {\hbar \ln(3) \over 8\pi M}=   4\ln(3) l_{\Pl}^2~.
\eneq 
This is the same spacing that one finds in the Schwarzschild black hole area spectrum according to Hod's interpretation.  Also, Maggiore's interpretation can only give a unique value for the spacing in the area spectrum if we use $\omega_I$ in the intermediate damping region, which gives  
\beeq
\Delta A_{total}= 32\pi M \hbar (|\omega_I|_n-|\omega_I|_{n-1})=32\pi M {\hbar \over 4 M}= 8\pi l_{\Pl}^2~.
\eneq 
This result is the same as the result that one finds for the Schwarzschild case using Maggiore's interpretation.

\vskip .5cm

\leftline{\bf Acknowledgments}
We are grateful to Emanuele Berti for clarifying an important issue for us regarding the Kerr spacetime and to Gabor Kunstatter for his feedback on the paper.


\def\jnl#1#2#3#4{{#1}{\bf #2} (#4) #3}

\def\RPP{{\em Reps.\ Prog.\ Phys. }}
\def\Zphys{{\em Z.\ Phys.} }
\def\jssc{{\em J.\ Solid State Chem.} }
\def\jpsJ{{\em J.\ Phys.\ Soc.\ Japan} }
\def\ptps{{\em Prog.\ Theoret.\ Phys.\ Suppl.\ } }
\def\PTP{{\em Prog.\ Theoret.\ Phys.\  }}
\def\LNC{{\em Lett.\ Nuovo.\ Cim.\  }}
\def\LRR{{\em Living \ Rev.\ Relative.} }
\def\JMP{{\em J. Math.\ Phys.} }
\def\NPB{{\em Nucl.\ Phys.} B}
\def\NP{{\em Nucl.\ Phys.} }
\def\PLB{{\em Phys.\ Lett.} B}
\def\PL{{\em Phys.\ Lett.} }
\def\PRL{\em Phys.\ Rev.\ Lett. }
\def\PRB{{\em Phys.\ Rev.} B}
\def\PRD{{\em Phys.\ Rev.} D}
\def\PR{{\em Phys.\ Rev.} }
\def\PRe{{\em Phys.\ Rep.} }
\def\AP{{\em Ann.\ Phys.\ (N.Y.)} }
\def\RMP{{\em Rev.\ Mod.\ Phys.} }
\def\ZPC{{\em Z.\ Phys.} C}
\def\SCI{\em Science}
\def\CMP{\em Comm.\ Math.\ Phys. }
\def\MPLA{{\em Mod.\ Phys.\ Lett.} A}
\def\IJMPB{{\em Int.\ J.\ Mod.\ Phys.} B}
\def\cmp{{\em Com.\ Math.\ Phys.}}
\def\JPA{{\em J.\  Phys.} A}
\def\CQG{\em Class.\ Quant.\ Grav.~}
\def\ATMP{\em Adv.\ Theoret.\ Math.\ Phys.~}
\def\PRSA{{\em Proc.\ Roy.\ Soc.\ Lond.} A }
\def\IJTP{\em Int.\ J.\ Theor.\ Phys.~}
\def\ibid{{\em ibid.} }
\vskip 1cm

\leftline{\bf References}

\renewenvironment{thebibliography}[1]
        {\begin{list}{[$\,$\arabic{enumi}$\,$]}  
        {\usecounter{enumi}\setlength{\parsep}{0pt}
         \setlength{\itemsep}{0pt}  \renewcommand{\baselinestretch}{1.2}
         \settowidth
        {\labelwidth}{#1 ~ ~}\sloppy}}{\end{list}}


\end{document}